# Large enhancement of infrared absorption due to trimer comprised of doping-N and S-S divacancies in the imperfect monolayer MoS$_2$: A first-principles study

Aijun Hong[*1]

[1]*Jiangxi Key Laboratory of Nanomaterials and Sensors, School of Physics, Communication and Electronics, Jiangxi Normal University, Nanchang 330022, China*

Correspondence and requests for materials should be addressed to A.J.H. (6312886haj@163.com) .

[**Abstract**]

In this study, we systematically study on crystal and electronic structures and optical absorption properties of perfect monolayer MoS$_2$ (M), M with S vacancy (M@SV), M with N doping at S site (M@ND) and M with both S vacancy and N doping at S site (M@V-D) using first-principles method. It is showed that the N atom is tend to located between Mo and S layers, leaving one vacancy at original site, to form interstitial N atom. Thus, the interstitial N atom and the S vacancy make up the N$_I$-V$_S$ dimer. We study M@V-D with five atomic configurations and find the most stable structure having the N$_I$-V$_S$-V$_S$ trimer. It is showed that the absorbance for the stable M@V-D in the most infrared region is obviously higher than that for the other systems. It is revealed that large enhancement of infrared absorption for the stable M@V-D is mainly attributable to the special electronic structure determined by the crystal structure with the trimer. It is considered that M@V-D could be the promising candidate for infrared materials.

# I. Introduction

In the past few years, two-dimensional (2D) materials such as grapheme [1], black phosphorus [2, 3] and $MoS_2$ [4-7] have received extensive attentions [8] due to enormous potential applications in sensor [9], spin [10], electrical, and optical devices [11]. A lot of meaningful theoretical work [12-15] on the monolayer $MoS_2$ is reported. For instance the S vacancy can increase the total magnetic moment of Mn-doped monolayer $MoS_2$ [16]. The monolayer $MoS_2$ with Cl doping at the S site is suitable for spin injection [17]. The Nb- and F-codoped monolayer $MoS_2$ has photocatalytic effect [18]. Some interesting experimental work [19-21] has also been reported. For examples the photoluminescence (PL) intensity of the monolayer $MoS_2$ is drastically improved via a molecular chemical doping technique or defect engineering and oxygen bonding [11]. The monolayer $MoS_2$ with a thickness of <1 nm can absorb up to 5−10% visible light [22]. These motive us to explore the effect of doping and vacancy on infrared absorption. Previous experimental and theoretical results [22-25] indeed show that visible light absorption coefficient (AC) of monolayer $MoS_2$ is interestingly large. However, it is almost zero in the infrared region. Therefore, it is almost impossible that the perfect monolayer $MoS_2$ is applied in the infrared sensing and detection applications. If increasing its absorption in the infrared region is achieved, this not only broadens the scope of application of the monolayer $MoS_2$, but also is beneficial to reveal the basic physical source.

Given the above considerations, we use the first-principles method to theoretically investigate crystal and electronic structures and optical absorption properties for perfect monolayer $MoS_2$, monolayer $MoS_2$ with S vacancy, monolayer $MoS_2$ with N doping at S site and monolayer $MoS_2$ with S vacancy and N doping at S site. For the sake of simplicity, we noted the above four systems respectively by M, M@SV, M@ND and M@V-D in the following statement. There are many possible atomic configurations for M@V-D due to the coexisting of N doing and S vacancy. In the letter, five atomic configurations (see Fig. 1) for M@V-D are considered and noted by

M@V1-D, M@V2-D, M@V3-D, M@V4-D and M@V5-D. It is found that there are defect and doping states respectively existing in M@SV and M@ND, and doping and defect states coexisting in M@V-D. M@V5-D has the most stable crystal structure and the most outstanding optical AC in the infrared spectrum. This is attributed to the special electronic structure determined by the crystal structure with trimer. In a word, it is revealed S vacancy and N doping coexisting improves not only the stability of structure but also the optical absorption. We consider that M@V-D is the promising candidate for infrared materials.

**II Computational details and methods**

All density functional theory (DFT) calculations on the perfect and imperfect $MoS_2$ monolayers are performed with Perdew-Burke-Ernzerhof (PBE) functional [26] and the cutoff energy 400 eV in the Vienna abinitio simulation package (VASP) code [26-29]. First, each structure of the four systems M, M@SV, M@ND and M@V-D with five atomic configurations is built on the three different supercells 3×3×1, 4×4×1 and 5×5×1 in order to study the effect of different doping and vacancy concentrations, namely impurity concentrations, on structures, electronic and optical properties. A 15 Å thickness of additional vacuum layer perpendicular to the *c* axis is inserted into each structure for avoiding the interactions caused by periodic images.

Second, the initial supercell structures are optimizes using the total energy convergence of $10^{-4}$ eV, and the maximum force on each atom is less than 0.05 eV/Å. For the calculations of structure optimization and electronic structure, the Monkhorst-Pack mesh is set to 5×5×1 ***k***-points. However, in order to obtain more accurate data in optical property calculation, the denser 21×21×1 ***k***-points is set.

Third, the cohesive energy $E_c$, indicating the stability of the solid material, is calculated using the formula

$$E_c = \sum_{j=1}^{N} n_j E_j - E_t \tag{1}$$

$N$ and $n_j$ represent the number of atomic species and the number of the kind of atoms $j$ in the unit cell. $E_j$ and $E_t$ are the energy of one isolated atom $j$ and the total energy for the unit cell. We also calculated the formation energy $E$ given by [30]

$$E = E_i - (E_p - E_V + E_D) \qquad (2)$$

where $E_i$ and $E_p$ respectively stand for the total energies of imperfect and perfect unit cells and $E_V$ is the energy of isolated S atom which is away from perfect unit cell and leads to the formation of vacancy, and $E_D$ is the difference between the energies of isolated doping atom and isolated atom replaced by doping atom.

Finally, the liner absorption coefficient $\alpha(\omega)$ and the absorbance $A(\omega)$ are studies in detail. It is well known that the $\alpha(\omega)$ and $A(\omega)$ can be described by the dielectric function $\varepsilon(\omega)$ which is the complex function

$$\varepsilon(\omega) = \varepsilon_1(\omega) + i\varepsilon_2(\omega) \qquad (3)$$

where $\omega$ is the angular frequency, $\varepsilon_1$, $\varepsilon_2$ are the real and imaginary parts, respectively. The liner absorption coefficient $\alpha(\omega)$ is given by the following equation:[31]

$$\alpha = \frac{\sqrt{2}\omega}{c}\left\{\left[\varepsilon_1^2 + \varepsilon_2^2\right]^{1/2} - \varepsilon_1\right\}^{1/2} \qquad (4)$$

where $c$ is a constant and is equal to the speed of light. In the optical properties, another more commonly used physical quantity is the absorbance [22]

$$A(\omega) = \frac{\omega}{c}\varepsilon_2 \Delta z \qquad (5)$$

here, $\Delta z$ is the thickness of the monolayer $MoS_2$. While the $\Delta z$ tends to 0, we can obtain another formula of the absorbance according to Taylor expansion [22]

$$A(\omega) = 1 - e^{-\omega\varepsilon_2 \Delta z / c} \qquad (6)$$

### III Results and discussion

**A. Crystal structures**

First, the influence of N doping at S site ($N_S$) or S vacancy ($V_S$) or $N_S$-$V_S$ coexisting in the crystal structure is studied. Five atomic configurations (see Fig. 1(c)) for M@V-D are considered and noted by M@V1-D, M@V2-D, M@V3-D, M@V4-D and M@V5-D, respectively. When modeling the N atom is fixed on the top S layer, and then the S vacancy has five possible sites noted by 1, 2, 3, 4 and 5. Sites of 1, 2 and the N atom are on the same S layer. Sites of 3, 4 and 5 are on the other S layer. The sites of 1, 2 and N are corresponding to the sites of 3, 4 and 5 on the other S layer. It is found that M@V5-D has the lowest total energy. Thus, only the lattice constants of M@V5-D are added to Fig. 1 (g)-(i). The single $V_S$ has a bigger effect on lattice constant than the single $N_S$ and the $N_S$-$V_S$ coexisting. It is seen that the lattice constants of M@SV, M@ND and M@V5-D all have obvious reduction. The reduction becomes bigger as the impurity concentrations increasing. The optimized crystal structures of M@SV, M@ND and M@V5-D are showed in Fig. 1 (a)-(f). It is seen that the $N_S$ leaves the site of replaced S atom, forming one S vacancy, to be located between Mo and S layers and becomes interstitial N atom. Thus, the N atom and the S vacancy make up $V_S$-$N_I$ dimer. In M@V5-D one $V_S$-$N_I$ dimer and one S vacancy make up a $N_I$-$V_S$-$V_S$ trimer when another S vacancy is introduced.

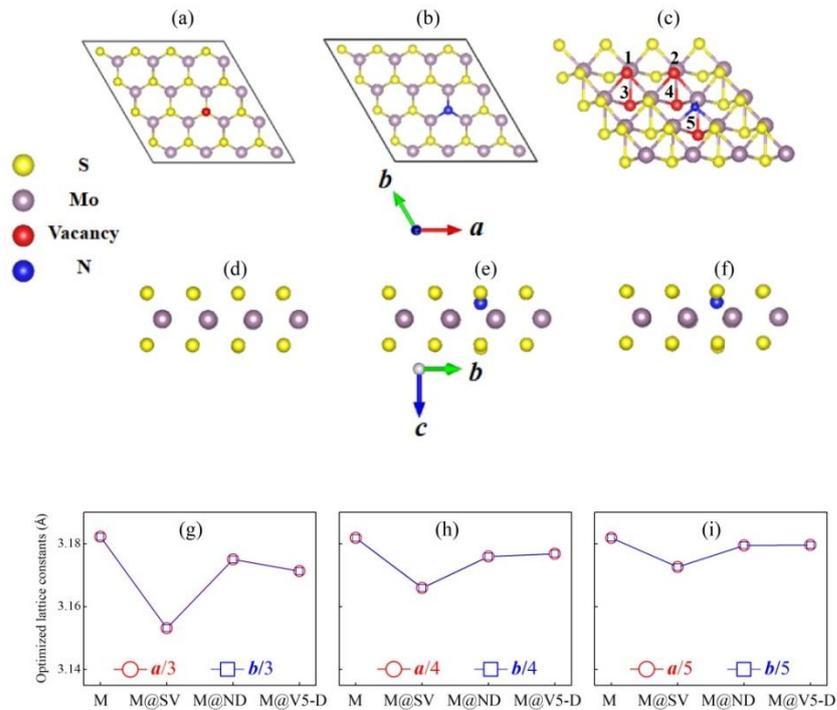

Fig. 1 Structural diagrams of optimized M@SV (a), M@ND (b), M@V-D (c) with 4×4×1 supercell from top view, and sequential (d)-(f) from side view. Optimized lattice constants of all systems with 3×3×1 (g), 4×4×1 (h) and 5×5×1 (i) supercell structures.

It is known that the stability of crystal structure can be evaluated by the magnitude of cohesive energy per atom $E_{CP}$. The bigger the $E_{CP}$ is, the more stable the structure is. The cohesive energy per atom $E_{CP}$ of each structure can be calculated using Eq. (1) and is listed in Table 1. It is indicated that $E_{CP}$ of M@V5-D at 3×3×1 supercell is the highest up to 7.348 eV, all five atomic configurations for M@V-D have higher $E_{CP}$ than the other three systems at the same size of supercell. The $E_{CP}$ of all systems increases with the impurity concentration increasing. $E_{CP}$ of M@V5-D is the most sensitive to impurity concentration and changes 0.09eV from 3×3×1 supercell to 5×5×1 supercell. It is worth to note that the perfect M possess the lowest $E_{CP}$ of 7.211eV, implying that the perfect M has the most unstable structure.

Table 1 Cohesive energy per atom $E_{CP}$ of each structure with different sizes of supercells (corresponding to different impurity concentrations).

| Supercell | Cohesive energy per atom $E_{CP}$ (eV) | | | | | | | |
|---|---|---|---|---|---|---|---|---|
| | M | M@SV | M@ND | M@V-D | | | | |
| | | | | M@V1-D | M@V2-D | M@V3-D | M@V4-D | M@V5-D |
| 3×3×1 | 7.211 | 7.231 | 7.296 | 7.317 | 7.326 | 7.321 | 7.322 | 7.348 |
| 4×4×1 | 7.211 | 7.222 | 7.255 | 7.269 | 7.271 | 7.271 | 7.271 | 7.285 |
| 5×5×1 | 7.211 | 7.218 | 7.241 | 7.249 | 7.250 | 7.249 | 7.249 | 7.258 |

Table 2 Formation energy $E$ of each structure with different sizes of supercells (corresponding to different impurity concentrations).

| Supercell | Formation energy $E$ (eV) | | | | | | |
|---|---|---|---|---|---|---|---|
| | M@SV | M@ND | M@V-D | | | | |
| | | | M@V1-D | M@V2-D | M@V3-D | M@V4-D | M@V5-D |
| 3×3×1 | 6.677 | -2.287 | 4.449 | 4.219 | 4.360 | 4.321 | 3.634 |
| 4×4×1 | 6.681 | -2.268 | 4.383 | 4.294 | 4.356 | 4.381 | 3.698 |
| 5×5×1 | 6.696 | -2.267 | 4.400 | 4.297 | 4.371 | 4.391 | 3.705 |

The formation energy $E$ can indicate the degree of difficulty for formation of material. The smaller the formation energy is, the easier the formation of material is. $E$ of each imperfect structure is also calculated by Eq. (2) and listed in Table 2. The $E$ for M@ND is negative and these for the other systems are positive. This implies that

M@ND is the easiest to form in all systems, M@V-D is medially difficult to form due to introduction of N dopant, while M@SV is the most difficult to form. It is well known that the monolayer MoS$_2$ with intrinsic S vacancies was successfully prepared. Thus, experimental preparation for M@ND and M@V-D is operable. *E* of both M@SV and M@ND is decreasing with impurity concentration increasing, while *E* and impurity concentration dependence in M@V-D is determined by relative positions of N atom and S vacancy. It is obviously seen that *E* of M@V2-D, M@V4-D and M@V5-D is decreasing with impurity concentration increasing while that of M@V1-D and M@V3-D is first decreasing and then increasing with impurity concentration increasing. The *E* difference between M@V1-D and M@V2-D is small as well as that between M@V3-D and M@V4-D at the same impurity concentration. *E* of M@V3-D at different 3×3×1, 4×4×1 and 5×5×1 supecells is 3.634, 3.698 and 3.705 eV, which is obviously lower than that of the other four atomic configurations for M@V-D. The *E* difference between M@V1-D and M@V5-D at the same 3×3×1 supercell is the largest up to 0.815 eV, and then the smallest *E* difference also reaches to 0.585 eV between M@V2-D and M@V5-D at the same 3×3×1 supercell. It is implied that M@V5-D is very stable and difficult to turn into the other atomic configurations due to thermal vibration of the lattice because thermal vibration induces the energy difference about $3k_BT = 0.078$eV which is far lower than 0.585 eV.

**B. Electronic structures**

In this section, the band structure and density of states (DOS) of each system at 3×3×1 supercell are presented in Figs. 2, and are discussed in detail. The calculated band structure for the perfect M at 3×3×1 supercell (see Fig. 2(a)) has the direct band gap of 1.64eV due to conduction band minimum (CBM) and valley band maximum (VBM) both located at the Γ point. As expected that the widths of band gaps at 5×5×1 and 4×4×1supercells are equal to that at 3×3×1supercell. However, the sites of CBM and VBM at 5×5×1 and 4×4×1supercells are both located at the K point. In fact, this is not contradictory, because the site of K point in the 3×3×1 supercell structure has folded to the site of Γ point. It is found in Fig. 2(a1) that the CBM originates from mainly Mo-d states and the VBM originates from hybridization between Mo-d and S-p states.

Fig 2 (b) shows defect states, consisted of two bands and closing to the CBM, are found inside the band gap in M@SV. Obviously, the defect states are mainly attributed to hybridization between Mo-d and S-p states. There is also small contribution from Mo-p states to the defect states. As expected in Fig 2 (c), the doping states appear in M@ND, which is only consisted of one band near the Fermi level. The doping states are mainly consisted of N-p, Mo-d and S-p states. The Mo-p states have small contribution to the doping states. The doping states in M@ND is close to the VBM, on the contrary, the defect states in M@SV is far from the VBM and is close to the CBM, suggesting that single S vacancy and single N doping produce p-type and n-type carriers, respectively.

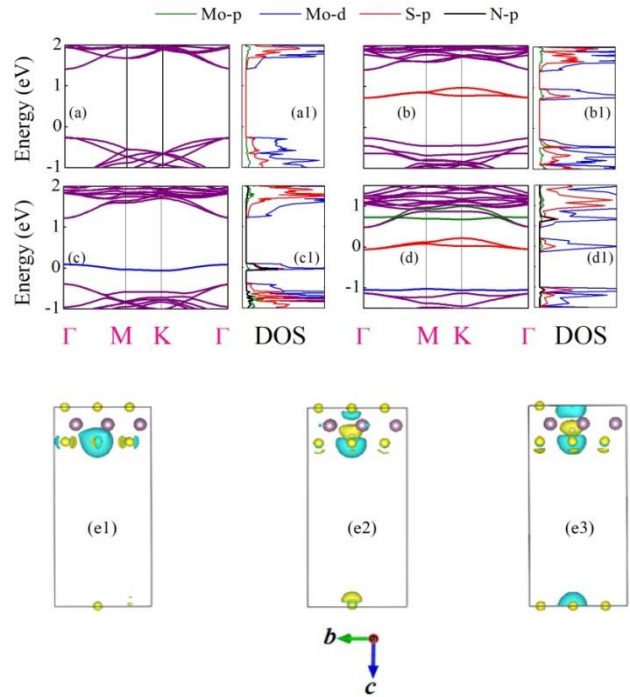

Fig. 2 Calculated band structures of M (a), M@SV (b), M@ND(c), M@V5-D (d) with 3×3×1 supercell. Corresponding atom-projected DOS of them is plotted in (a1), (b1), (c1) and (d1). Three dimensional charge density difference of M@SV (e1), M@ND (e2), M@V-D (e3) with 3×3×1 supercell, yellow and blue mean to gain and lose electrons, respectively.

The band structure of M@V5-D in Fig. 2(d) shows that doping states and defect states coexists in the band gap. Compared with the other three systems, the Fermi level of M@V5-D is closer to the CBM, and crosses the doping states. The defect states are close to the VBM, and are far below the Fermi level while the Fermi level crosses the defect states in M@ND. Interestingly, the N-p states have small contribution to other two conduction bands in M@V5-D (plotted by the olive lines in Fig.2 (d)).

The band structure and density of states (DOS) of each system with 4×4×1 and 5×5×1 supercells are plotted in Fig. 3. Defeat states retain to be in M@SV as well doping states are in M@ND. Defeat states and doping states coexists in M@V5-D, also. The defect and doping states become flat and degenerate with the size of supercell increasing.

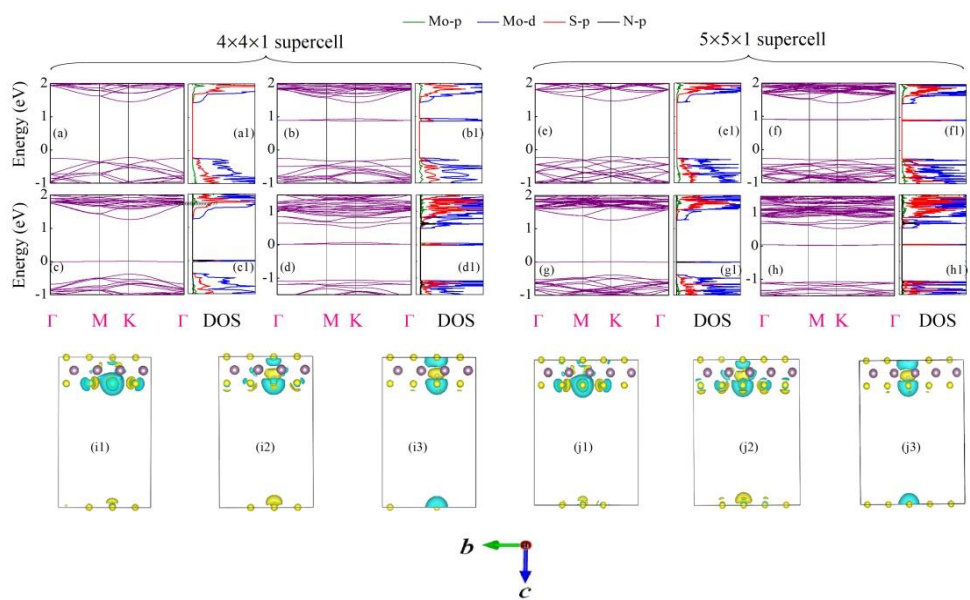

Fig. 3 Calculated band structures of M (a), M@SV (b), M@ND(c), M@V5-D (d) with 4×4×1 supercell and sequential (e)-(h) with 5×5×1 supercell. Corresponding atom-projected DOS of them with 4×4×1 supercell is plotted in (a1), (b1), (c1) and (d1), these with 4×4×1 supercell is plotted in (e1), (f1), (g1) and (h1). Three dimensional charge density difference of M@SV (i1), M@ND (i2), M@V-D (i3) with 4×4×1 supercell. These with 5×5×1 supercell are showed in (j1), (j2), (j3) and (j4). Yellow and blue mean to gain and lose electrons, respectively.

In order to study the effect of doping and vacancy on charge transfer, the charge density difference (CDD) is defined as the difference between the charge densities of imperfect and perfect M, and the CDDs for M@SV, M@ND and M@V-D with 3×3×1 supercell are calculated and then presented in Fig. 2 (e1)-(e3). Those with the other sizes of supercell are added to Fig. 3 (i1)-(i3) and (j1)-(j3) in which the yellow and blue separately denote charge increase and charge decrease. It is found that S vacancy has great effect on the charge transfer of the S atom layer where the S vacancy is, and has slight influence on that of the other S layer. Fig. 1(e1), Fig. 2(i1) and (j1) all show that the charge transfers from S vacancy to the vicinity of S atoms surrounding S vacancy. For N doping, charge transfer occurs in both two S layers and charge accumulates around the N atom from two S layers. For N doping and S vacancy coexisting, charge accumulates around the N atom mainly from two S vacancies.

## C. Optical properties

It is known that the dielectric function is a tensor composed of nine components. Thus, the absorption coefficient is also a tensor with nine components. We put focus on optical properties along the layer–normal direction, and plot the absorption coefficient and the absorbance in Fig. 4.

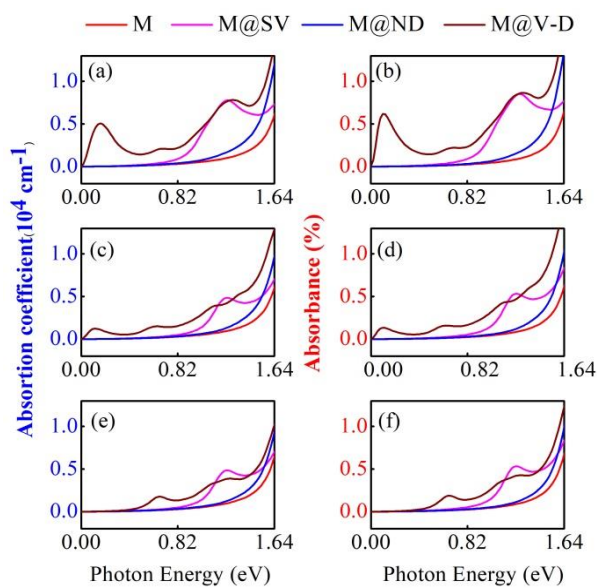

Fig. 4 Calculated absorption coefficients of all systems with different 3×3×1 (a), 4×4×1 (c), and 5×5×1 (e) supercells, calculated absorbances of all systems at different 3×3×1 (b), 4×4×1 (d), and 5×5×1 (e) supercells.

The absorption coefficient and absorbance are two different physical quantities; they have a corresponding relationship according to Eq. (6). In this letter, we pay attention to discuss the absorbance which is dimensionless and equal to the ratio of energy flux densities for absorbed and incident light. Fig. 4 (b), (d) and (f) shows that the absorbances for the M and M@ND with all the sizes of supercell is fairly small in the most energy range of 0.0012eV to 1.64eV (infrared region), and increase with photon energy increasing. It is worth to mention that there no absorption peak in the absorbances for M and M@ND. However, the absorbance for the M@SV has obvious absorption peak near 1.2eV ascribed to the defect states existing inside the band gap. The absorption peak is sensitive to the size of supercell, which change from 0.7% to 0.3% when 3×3×1 supercell is expanded to 5×5×1 supercell. Nevertheless, the peak position is barely affected by the size of supercell. There is obvious absorption peak at 0.1 eV in the M@V-D with 3×3×1 supercell. The peak values decrease with the impurity concentration increasing. There are the other two peaks in M@V-D respectively located at 0.67 and 1.2 eV. The former peak becomes obvious while the later becomes disappeared as the impurity concentration increases.

**IV Summary**

In the letter, we reported that the crystal and electronic structures and optical properties of the perfect monolayer $MoS_2$ and three kinds of imperfect monolayer $MoS_2$ with different impurity concentrations. It is found that doping N at the site of S atom tends to form an interstitial N atom leaving the vacancy, the N atom and the vacancy constitute the $N_I$-$V_S$ dimer. Thus, there is $N_I$-$V_S$-$V_S$ trimer in monolayer $MoS_2$ with doping N and S vacancy. Doping states and defect states respectively appears in the band gap of M@ND and M@SV, and both coexist in the band gap of M@V-D. It is revealed the hybridization of Mo-4d and S-3p orbitals yield the defect states, and the mutual hybridizations of Mo-4d, S-3p and N-2p orbitals yield the doping states. We predicted that the infrared absorbance of the monolayer $MoS_2$ with S vacancy and doping N coexisting is the most excellent. It is revealed that special electronic structure determined by special crystal structure of M@V-D with $N_I$-$V_S$-$V_S$ trimer.

**Acknowledgement:**

This work is supported by the National Natural Science Foundation of China (Grant No. 11804132), the Science and Technology Research Project of Department of Education of Jiangxi province (Grant No. 160340), and the Open Fund of Jiangxi Key Laboratory of Nanomaterials and Sensors(Grant No. 1700001).